\documentclass[usenatbib]{mn2e}
\usepackage{graphicx}
\usepackage{subfigure}
\usepackage[colorlinks=true,linkcolor=black,citecolor=black,urlcolor=blue]{hyperref}
\usepackage{aas_macros}
\usepackage{enumerate}
\usepackage{amsmath}
\usepackage{mathtools}
\usepackage{IEEEtrantools}
\usepackage{float}
\usepackage{txfonts}
\usepackage[applemac]{inputenc}
\usepackage[english]{babel}
\usepackage{booktabs}
\usepackage{multirow}
\usepackage{amstext}
\usepackage{subfigure}
\usepackage{graphicx}
\usepackage[np,noautolanguage]{numprint}
\usepackage{url}
\usepackage{hyperref}
\usepackage{pdfsync}
\usepackage{threeparttable}
\usepackage{color}

\newcommand{\Msun}{M$_{\odot} $}
\definecolor{grey}{rgb}{0.5,0.6,0.7}
\def \simlt { \lower .75ex \hbox{$\sim$} \llap{\raise .27ex \hbox{$<$}} }
\definecolor{purple}{rgb}{0.65,0.15,0.9}
\definecolor{darkorange}{rgb}{0.8,0.3,0}
\definecolor{olive}{rgb}{0.4,0.6,0.25}
\definecolor{darkgreen}{rgb}{0,0.7,0}
\definecolor{darkred}{rgb}{0.5,0,0}

\title[Black holes and non-Gaussianities]{Black hole formation and growth with non-Gaussian primordial density perturbations}
\author[Habouzit et al.]{M\'{e}lanie Habouzit$^{1}$\thanks{E-mail: habouzit@iap.fr},
 Marta Volonteri$^{1}$,
  Muhammad Latif$^{1}$,
  Takahiro Nishimichi$^{2,3}$,
  \newauthor
   S\'{e}bastien Peirani$^{1}$,
    Yohan Dubois$^{1}$,
     Gary A. Mamon$^{1}$,
      Joseph Silk$^{1,4}$
       and Jacopo Chevallard$^{5}$\\
$^1$Institut d'Astrophysique de Paris, Sorbonne Universit\'{e}s, UPMC Univ Paris 6 et CNRS, UMR 7095, 98 bis bd Arago, 75014 Paris, France\\
$^2$Kavli Institute for the Physics and Mathematics of the Universe, The University of Tokyo Institutes for Advanced Study, 5-1-5 Kashiwanoha, Kashiwa 277-8583, Japan\\
$^3$CREST, JST, 4-1-8 Honcho, Kawaguchi, Saitama, 332-0012, Japan\\
$^4$Department of Physics and Astronomy, The Johns Hopkins University
Homewood Campus, Baltimore MD 21218, USA\\
$^5$Scientific Support Office, Directorate of Science and Robotic Exploration, ESA/ESTEC, Keplerlaan 1, 2201 AZ Noordwijk, The Netherlands}

\begin{document}
\maketitle

\begin{abstract}
Quasars powered by massive black holes (BHs) with mass estimates above a billion solar masses have been identified at redshift 6 and beyond. The existence of such BHs requires almost continuous growth at the Eddington limit for their whole lifetime, of order of one billion years. In this paper, we explore the possibility that positively skewed scale-dependent non-Gaussian primordial fluctuations may ease the assembly of  massive BHs. In particular, they produce more low-mass halos at high redshift, thus altering the production of metals and ultra-violet flux, believed to be important factors in BH formation. Additionally, a higher number of progenitors and of nearly equal-mass halo mergers would boost the mass increase provided by BH-BH mergers and merger-driven accretion. We use a set of two cosmological simulations, with either Gaussian or scale-dependent non-Gaussian primordial fluctuations to perform a proof-of-concept experiment to estimate how BH formation and growth are altered. We estimate the BH number density and the  fraction of halos where BHs form, for both simulations and for two popular scenarios of BH formation (remnants of the first generation of stars and direct collapse in the absence of metals and molecular hydrogen). We find that the fractions of halos where BHs form are almost identical, but that non-Gaussian primordial perturbations increase the total number density of BHs for both BH formation scenarios by a factor of two. 
We also evolve BHs using merger trees extracted from the simulations and find that both the mean BH mass and the number of the most massive BHs at $z=6.5$ are 
up to twice the values expected for Gaussian primordial density fluctuations.

\end{abstract}


\section{Introduction}
\label{sec:intro}
Primordial density perturbations evolve with time, cause the collapse of dark matter halos and lead to the formation of large scale structures. As the hot big bang theory has no explanation for the distribution of these density fluctuations, inflation  has been considered to be a natural physical process able to produce the necessary spectrum of the density perturbations. The simplest inflationary models, a single scalar field slowly rolling down a shallow potential, predict a very nearly Gaussian distribution of these density fluctuations  \citep{Gangui94,Acquaviva2002,Maldacena03}. Primordial density perturbations place the tightest constraints on inflationary models and on how physical processes at very high energies shaped the Universe at very early times.\\
Primordial perturbations described by a Gaussian distribution are supported, on large scales, by measurements of the temperature anisotropies of the cosmic microwave background (CMB), which are the relics of density perturbations in the cosmic fluid at the time of last scattering.
Planck's results \citep{PlanckCollaboration+13_cosmopars} have made  incomparable progresses in the accuracy of the estimated cosmological parameters and on our knowledge of the beginning of the Universe. The Planck mission has, however, focused on large structures, considering primordial density perturbations on the scale of clusters. By mapping in detail the Cosmic Microwave Background (CMB) on the full sky, Planck has provided very strong constraints on the local non-Gaussianities \citep{Gangui94} by estimating the parameter describing the quadratic coupling of the primordial perturbations (introduced by \cite{Komatsu+01}), $f_{\rm{NL}}=2.7 +/- 5.8$ \citep{PlanckCollaboration+13}.  
However, as predicted by some inflationary models, non-Gaussianities on smaller scales, beyond the reach of Planck measurements, are still conceivable. \\
Recent studies have shown that scale-dependent non-Gaussianities, consistent with Planck constraints at large scale, can have an important impact on structure formation on galactic scales.
\citet{Habouzit2014} used cosmological dark matter simulations to investigate the impact of scale-dependent non-Gaussian primordial perturbations, predicted by some inflationary models (Alishahiha, Silverstein \& Tong 2004; Silverstein \& Tong 2004; Chen 2005). They compared 5 simulations: a Gaussian simulation and 4 simulations based on scale-dependent non-Gaussian prescriptions for the initial conditions (all consistent with Planck's constraints). The non-Gaussian initial models developed of \citet{Habouzit2014}  employed a low level of non-Gaussianities on scales of galaxy clusters and larger ($\rm{log(f_{NL}^{local})<1}$ for $\rm{log(k/Mpc^{-1})<0.75}$, to be consistent with Planck's results) and a higher level on smaller scales ($\rm{log(f_{NL}^{local})>1}$ for $\rm{log(k/Mpc^{-1})<-0.5}$). Applying a galaxy formation model using the redshift-dependent stellar-halo mass relation of \citep{Behroozi+13} to paint galaxies on dark matter halos, \citet{Habouzit2014}  find that, with non-Gaussian initial conditions, there is a significant enhancement (up to 0.3 dex at redshift $\ge$ 10) of the halo and galaxy mass function, which increases with redshift and decreases with halo/galaxy mass.  The galaxy mass function is significantly altered when non-Gaussianity varies strongly with scale.\\
Using the same set of simulations, \citet{Chevallard2014} went further to address the implications of scale-dependent non-Gaussianities on cosmic reionization. They considered a modified semi-analytical galaxy formation model based on \citet{Mutch2013} to compute the stellar mass assembly in each dark matter halo, and used the \citet{Bruzual2003} stellar population synthesis code to compute the far-UV luminosity function for different redshifts. Reionization is thought to be mainly driven by UV radiation emitted by massive stars born in the first galaxies because hydrogen ionizing photons can escape more easily from low mass halos than high mass ones. The number of ionizing photons emitted by early galaxies depends on their number density,  i.e. on their far-UV galaxy luminosity function. \citet{Chevallard2014} employed different reionization models (a fixed escape fraction $f_{\rm{esc}}=0.2$, and two different escape fractions varying with redshift) to investigate the ionization fraction of the Universe as a function of redshift, and concluded that in the most favorable case (strongest non-Gaussian model), the Universe can be reionized earlier, in better agreement with the electron Thomson scattering optical depth  measured by Planck.
\\ 
The population of  BHs powering quasars at $z>6$~\citep{fanetal06, jiangetal09, mortlocketal11}  represent an issue similar to that of reionization. In the studies of reionization, for theory to match the measured electron Thomson scattering optical depth, it is necessary to assume that the escape fraction is $f_{\rm{esc}}\sim0.2$, much larger than observed for the typical galaxy. In a similar way, in the case of BHs, for theory to match their masses and number densities, it is necessary to assume that BHs grow almost continuously at the Eddington limit (or continuously at almost  the Eddington rate) for a billion years: a MBH with initial mass $M_0$ grows with time $t$ as $M=M_0\exp\left\{\left[({1-\eta})/{\epsilon}\right]({ t}/{t_{\rm{Edd}}})\right\}$, 
where $t_{\rm{Edd}}={\sigma_T \,c}/({4\pi \,G\,m_p})=0.45\,{\rm Gyr}$,  $\eta$ is the fraction of rest mass energy released by accretion, and $\epsilon \leqslant \eta$ the radiative efficiency. In thin accretion disks \citep{Shakura73}, $\eta=\epsilon$, with $\epsilon$ ranging from 0.057 to 0.32 for BH spin ranging from 0 to 0.998.
The BH masses of $z>6$ quasars can reach $ 10^{10}$ \Msun \citep{Wu2015}, therefore constant Eddington-limited accretion for the whole Hubble time is implied if $M_0<10^2$ \Msun $\,$ and $\epsilon\sim 0.1$.
While this assumption is not impracticable, it stretches the typical properties of BHs and quasars. It is therefore worthwhile to assess whether scale-dependent non-Gaussianities can increase BH growth as they boost reionization powered by galaxies. 

The formation and growth of BHs in a cosmology including non-Gaussian primordial density fluctuations can be altered in several ways. In the first place, a larger number of galaxies may be able to form a BH. Most theories  \citep[for a review, see][and references therein]{2010A&ARv..18..279V} link BH formation to the first generation of galaxies, either via the first stars \citep[Pop III stars, stars without heavier elements than hydrogen and helium,][]{Madau2001,Volonteri2003}, via gas collapse in metal-free halos illuminated by strong photo-dissociating flux  \citep[`direct collapse, DC,][]{Loeb1994,Bromm2003,spaans06,Begelman2006,Dijkstra2008,2013MNRAS.433.1607L}, or via mergers of stars or stellar-mass BHs in dense stellar clusters \citep{Devecchi2009,2011ApJ...740L..42D}. The formation of BHs in the cosmology we propose could be impacted in two ways; the enhancement at the low-mass end of the galaxy mass function at high redshift could increase the number of halos producing stars, thus boosting the formation of BHs as Pop III remnants. Regarding the DC scenario, the number of  halos illuminated by dissociating radiation could also be enhanced because of the higher star formation. On the other hand, the enhanced stellar production would also lead to increased metal pollution, suppressing the ``eligibility" of a fraction of halos. Therefore some BH formation mechanisms would be boosted, and other may be suppressed in a way non-trivial to predict. \\
The growth of BHs is also impacted. There are two channels for BHs to grow in mass: the first one is by BH-BH mergers, the second one by accretion of gas, which can be strongly increased during galaxies merger episodes. Both channels are facilitated in the presence of non-Gaussianities, because of the increased number of low-mass galaxies, which increases the number of galaxy mergers, and of BH mergers as well \footnote{A small fraction of merging BHs may however be ejected from the host halos because of the gravitational wave induced recoil, \citep[e.g.,][]{Redmountrees,2007ApJ...667L.133S} thus lowering the `positive’ contribution of BH-BH mergers}. 

In the present paper, we compare the formation and growth of BHs in a Gaussian simulation and the most non-Gaussian simulation of \citet{Habouzit2014}, G and NG4 thereafter.  In Section 2, we recall the main features of the two simulations using initial conditions with either Gaussian or non-Gaussian primordial perturbations. 
In Sections 3 and 4, we investigate the formation of BHs through two formation scenarios: DC (Section 3) and the remnants of the first generation of stars (Section 4). 
We build a model to compute the Lyman-Werner radiation that may impinge on each halo in the two simulations. Looking at all halos evolving under a radiation higher than $J_{\rm{21,crit}}$, we are able to estimate where DC BHs can form for the two simulations.
Similarly, Section 4 is devoted to the study of the number density of BHs formed via the remnants of the first generation of stars.
In Section 5, we follow the most massive halos in both simulations with the aid of merger trees to perform an analysis addressing the growth of BHs over cosmic time.


\section{Simulation parameters}
\subsection{Initial conditions: prescription for scale-dependent non-Gaussianities}
\label{subsec:fnl}

We employ a simple model to have a significant amount of
non-Gaussianity on small scales, relevant for early structure formation,
while keeping them small on large scales to meet the strong constraints obtained by the
Planck CMB mission \citep{PlanckCollaboration+13}.
Namely, we investigate here the {\it generalized local ansatz} proposed
by \cite{Becker+11}:
\begin{equation} 
\phi(\mathbf{x}) = \phi_\rmn{G}(\mathbf{x}) + \left[f_\rmn{NL} *
  (\phi_\rmn{G}^2 - \langle{\phi_\rmn{G}^2}\rangle)\right](\mathbf{x}) 
\label{eq:localansatz}
\end{equation}
where $\phi(\mathbf{x})$ is the curvature perturbation,  $\phi_\rmn{G}$ a Gaussian random field and where the operation $(f_\rmn{NL} * A)$ is a convolution of a random variable $A$ and 
a $k$-dependent kernel defined in Fourier space:
\begin{equation} 
f_{\rmn{NL}}(k) = f_{\rmn{NL}, 0}\,\left(\frac{k}{k_0}\right)^{\alpha}.
\label{eq:nGmodel}
\end{equation}

We explore four different models by varying the normalization
$f_{\rm NL,0}$ and the slope $\alpha$ of $f_{\rm NL}(k)$, in such a way that
the non-Gaussianity is significant on galactic scales, yet small enough to
meet the current constraints from the Planck mission \citep{PlanckCollaboration+13}. We restrict ourselves to positively skewed primordial density fluctuations, i.e. $f_{\rm NL}>0$, hence $f_{\rm NL,0}>0$.

We modify the initial condition generator originally developed by \citet{Nishimichi+09},
based on second-order Lagrangian perturbation theory
\citep[e.g.,][]{Scoccimarro98,Crocce+06}, parallelized by \citet{Valageas+11}
and with local-type non-Gaussianities implemented by \citet{Nishimichi12}.
We follow \citet{Becker+11} and realize the generalized local ansatz of
equation (\ref{eq:localansatz}) by taking
a convolution of the curvature squared and the $k$-dependent $f_\mathrm{NL}$
kernel in Fourier space.  We use the public Boltzmann code, {\tt CAMB}
\citep{Lewis+00} to compute the transfer function and multiply it to the
curvature perturbations to have the linear density fluctuations.

\subsection{N-body simulations, halo catalog and merger trees}

We have performed five cosmological simulations with {\sc Gadget-2}
\citep{Springel05} for a $\Lambda$CDM
universe using Planck parameters \citep{PlanckCollaboration+13_cosmopars}, namely
$\Omega_{\rm{m}}=0.307$,  $\Omega_{\Lambda}=0.693$,  $h=0.678$, $\sigma_8=0.829$ and $n_{\rm{s}}=0.9611$.
Each simulation was performed in a periodic box of side $50\,h^{-1}$ Mpc
with $1024^3$ dark matter particles (e.g. with mass resolution
of $\sim 9.9 \times 10^6\,h^{-1}\,\rm M_\odot$).
One simulation uses Gaussian initial conditions (hereafter, `G'), while the others
consider non-Gaussian initial conditions (in this work, only the most powerful non-Gaussian simulation, `NG4' is used, eqs.~[\ref{eq:localansatz}] and
[\ref{eq:nGmodel}]). For simulation G, we use $f_{\rm NL,0}=0$, while simulation NG4  is described by $f_{\rm NL,0}=10^{4}$ and  $\alpha=4/3$ (see eq.~\ref{eq:nGmodel}), both simulations with the same initial phases.
The simulations started at $z=200$ and ended at $z=6.5$.
In each case, the Plummer-equivalent force softening adopted
is 5\% of the mean inter-particle distance
($2.44\,h^{-1}\,\rm kpc$), kept constant in comoving units.

 For each snapshot (taken every $\sim40\, \rmn{Myr}$), catalogues of
halos are prepared using {\sc AdaptaHOP} \citep*{Aubert+04}
which uses an SPH-like kernel to compute
densities at the location of each particle and partitions
the ensemble of particles into subhalos based on saddle points
in the density field.
The use of this code is then
attractive, since
subhalos are separated from their parent halos.
Note that, in our study, only halos or subhalos containing at least 20 particles  (i.e. $2.9\times
10^8 \rmn{M}_\odot$) were retained in the different catalogues.
We then study the individual evolution of (sub)halos, by
building halo merger trees using  {\sc TreeMaker} \citep{Tweed+09}.
This latter allows us to obtain the evolution of physical properties of each
(sub)halo, such as the mass  and the list of all of its progenitors.
Thus, one can derive accurately the evolution of the mass of each dark
matter (sub)halo.

\begin{figure}
	\centering
	\includegraphics[scale=0.5]{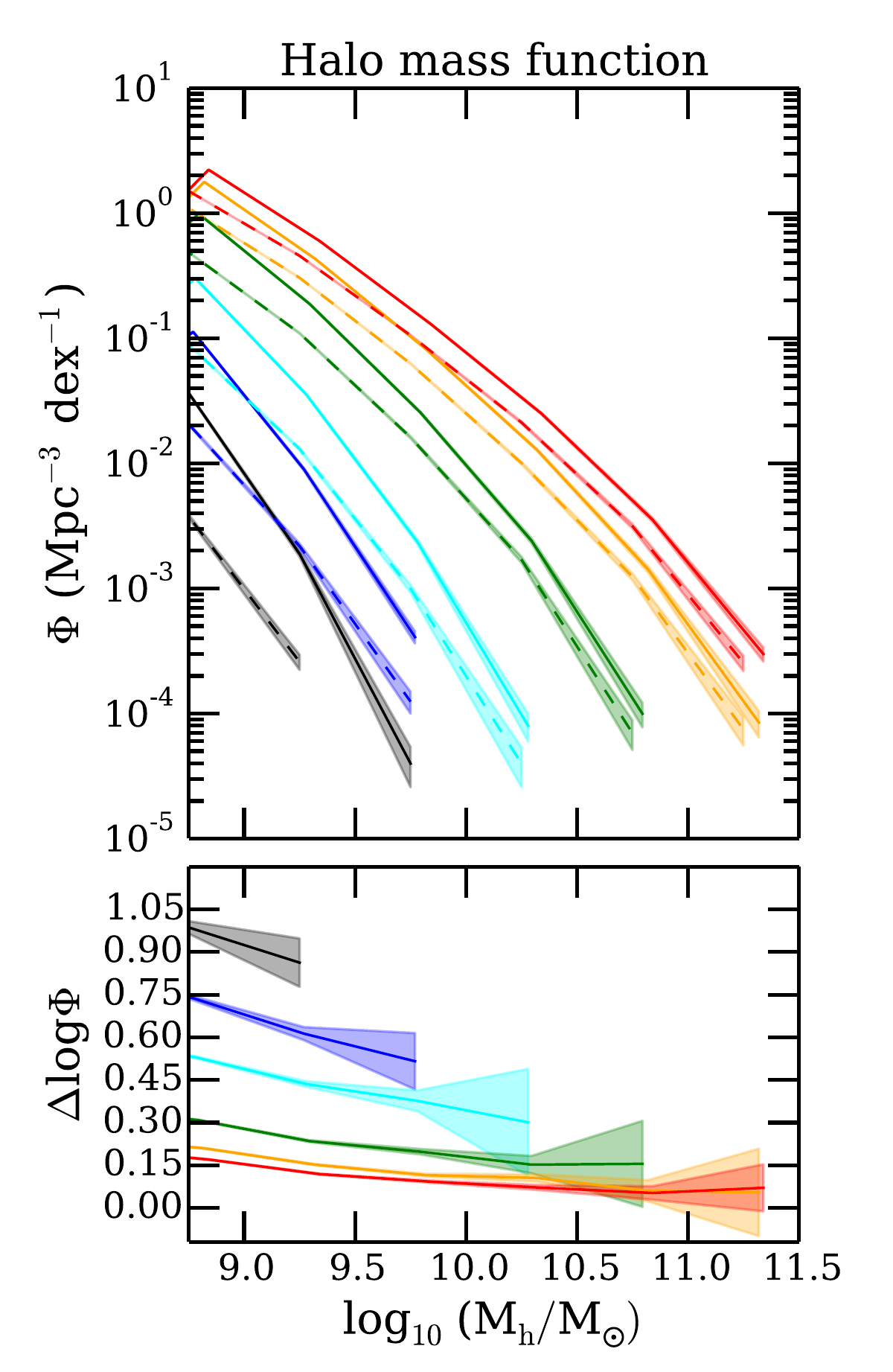}
	\caption{Halo mass functions (top panel) for the Gaussian G (dashed lines) and non-Gaussian NG4 (lines) simulations. Residual of the log mass function between Gaussian and non-Gaussian simulations is shown on the bottom panel. 
The different curves indicate different redshifts: $z=17$ (black), $z=15$ (blue), $z=13$ (cyan), $z=10$ (green), $z=8$ (orange), $z=7$ (red).}
	\label{fig:massfunctions}
\end{figure}

Fig ~\ref{fig:massfunctions} shows the halo mass function for the Gaussian (G, dashed lines) and the non-Gaussian (NG4, solid lines) simulations as a function of the virial halo mass, for different redshifts from z=17 to z=7 (see also Habouzit et al. 2014). The main conclusion is that primordial non-Gaussianities introduce an enhancement in the halo mass function, which increases with redshift and decreased with halo mass. At z=10 (green curves, bottom plot), this enhancement is up to 0.3 dex.
The consequences of positively skewed non-Gaussian initial conditions with a blue tilt (i.e. $\alpha>1$) are that the formation of less massive halos are amplified more significantly.
This is simply because the formation of these halos originate from initial fluctuations on small scales (i.e. large $k$) where we put more non-Gaussianity. 
Then, the mode transfer from large to small scales through the nonlinear nature of structure formation driven by gravitational instability gradually surpasses the initial signal on small scales, resulting in the suppression of the halo mass function ratio with time (see bottom panel of Fig ~\ref{fig:massfunctions}). At redshift z=7, the ratio is 0.15 dex or smaller. A similar halo abundance at the final time but with initially very different halo mass function naturally indicates that the merger history is different in the two simulations.
It is thus well-motivated to go beyond the simple halo abundance comparison and consider the evolution of BH mass in detail by following the merger history of each halo.
Since such investigation is difficult with analytical calculation, we here resort to numerical simulations and build a simple model for the formation and the evolution of BHs.
 
%


\section{Modeling the Lyman Werner radiation to estimate the distribution of potential DC BHs}
In this section, we study whether the probability of forming BHs  in the DC scenario \citep{Loeb1994,Bromm2003,Koushiappas2003,Begelman2006,Lodato2006,Dijkstra2008,2013MNRAS.433.1607L} is higher or lower when scale-dependent non-gaussianities produce more low-mass halos . The DC scenario is very appealing as it may lead to the formation of large $10^{4}-10^{6}$ \Msun $\,$seeds, that ease the growth constraints for the sample of $z>6$ quasars.\\
Metal-free halos at high redshift ($z=20$ and later) may host the formation of DC BHs under specific conditions. If the inflow rate of gas at the center of the halo is higher than $0.1 \rm{M}_{\odot}/\rm{yr}$, a supermassive star-like object \citep{Begelman2006,spaans06,Begelman2007b,Begelman2010,2011MNRAS.414.2751B} forms in the nucleus, and this star collapses and forms a  BH with mass up to 90\% of the stellar mass.  Metals, able to efficiently cool the gas to low temperatures, down to CMB temperature, would strongly decrease the Jeans mass, thus fostering fragmentation and star formation, decreasing the inflow rate needed to form only one massive object in the center of the gas cloud. The presence of molecular hydrogen could also cool the gas; a strong photo-dissociating radiation (Lyman-Werner, LW, photons $11.2\, \rm{eV}<E_{\rm{LW \,photons}}<13.6 \,\rm{eV}$) is then needed to destroy molecular hydrogen and prevent its formation. The conditions advocated for DC models are therefore: metal-free conditions and presence of a sufficiently strong dissociating LW flux.\\
The level of ``critical" LW radiation is a major factor in the DC BH scenario. Several aspects have been addressed in the last years. Despite the fact that the mean LW radiation background is orders of magnitude lower than the radiation intensity required to keep the fraction of molecular hydrogen close to zero ($<10^{-4}$), spatial variations in the LW intensity appear to be a key element in the DC mechanism \citep{Dijkstra2008}.
Using  simulations and an analytic framework, \citet{Shang2010}, \citet{Ahn2008}, \citet{Agarwal2012SAM}, and \citet{Agarwal2014} find that spatial variations in the LW radiation exist and are due to clustering of the LW photon sources and the  matter density fluctuations (large-scale structures):  the proximity with star-forming regions is  essential for the DC scenario.
These studies suggest that a halo can be exposed to a high enough LW radiation intensity if it lives in a clustered environment, close to star-forming galaxies.
\citet{Dijkstra2008} compute the probability distribution function of the LW radiation that irradiates halos at redshift $z=10$ and show that a small fraction of halos ($10^{-8} \, \rm{to} \, 10^{-6}$) can be exposed to a radiation higher  $J_{\rm{21,LW,crit}}\sim10^{3}$ (in units of $10^{-21}\, \rm{erg}\, \rm{s}^{-1}\, \rm{cm}^{-2}\, \rm{Hz}^{-1}\, \rm{sr}^{-1}$), a value shown by  \citet{Bromm2003} to lead to a sufficiently low molecular hydrogen fraction. \citet[][D14 hereafter]{Dijkstra2014} suggest that the number density of DC BHs is sufficient to explain that of $z>6$ quasars if $J_{\rm{21,LW,crit}}\sim10^{2}$.\\

In the meanwhile, \citet{Latif2015} use zoomed cosmological simulations of single halos to find that complete molecular hydrogen dissociation may not be necessary, while \citet{LatifXray2015} and \citet{2014arXiv1411.2590I}  include the impact of X-rays on molecular hydrogen dissociation and show that  X-rays make DC BHs  rarer than previously expected, less than  the number density of $\sim$ 1 Gpc$^{-3}$ necessary to explain the population of $z>6$ quasars. However, non-Gaussianities provide an enhancement in the low-mass end of the halo/galaxy mass function, therefore they can increase the probability of having halo/galaxy clustered regions, hence boosting the number density  of eligible DC regions in the early Universe.\\

\subsection{The model}
Our model is a modification of D14, where we adopt dark matter simulations to obtain the clustering of halos and their redshift evolution, rather than analytical prescriptions \citep[see][for a discussion of the uncertainties in clustering assumptions]{2014arXiv1411.2590I}. To identify halos which can potentially form a DC BH we use the LW radiation model of D14, described in the following.

\noindent
The stellar mass of a dark matter halo $M_{h}$ is assigned as:
\begin{equation}
M_{\rm{\star}}=f_{\rm{\star}} M_{\rm{h,gas}} = f_{\rm{\star}} \frac{\Omega_{\rm{b}}}{\Omega_{\rm{m}}} M_{\rm{h}},
\end{equation}
where $f_{\rm{\star}}=0.05$ is the fraction of gas which turns into stars, $M_{\rm{h,gas}}$ the gas mass of the halo,  $M_{\rm{h}}$ the total mass of the halo, $\Omega_{\rm{b}}$ the baryon density and $\Omega_{\rm{m}}$ the total matter density.
The mean production rate of LW photons per solar mass of star formation is time-dependent, where time is counted from the time $t_{\rm{Myr}}$ when a burst of star formation occurs, and expressed as
\begin{equation}
\left<Q_{\rm{LW}} (t) \right>= Q_{\rm{0}} \left(1+\frac{t_{\rm{Myr}}}{4}\right)^{-3/2} \exp \left({-\frac{t_{\rm{Myr}}}{300}}\right) \quad \rm{s^{-1}}\, \rm{M_{\rm{\odot}}^{-1}}.
\end{equation}
with $Q_{0}=10^{47}  \rm{M_{\odot}} \rm{s^{-1}}$.\\
The mean production rate is computed one free-fall time after the star formation burst. Assuming that $t_{\rm{ff}}=\sqrt{3 \pi/(32 \,G \, \rho)}=\sqrt{3 \pi/(32 \,G \,200\, \rho_{\rm{c}})}=\sqrt{3 \pi/(32 \,G \,200\, (1+z)^{3}\, \rho_{\rm{c,0}})}$, the free-fall time can be expressed as:
\begin{equation}
t_{\rm{Myr, ff}}\sim 83 \left( \frac{1+z}{11}\right)^{-3/2}.
\end{equation}
D14 motivate this choice by the requirement that the molecular hydrogen is suppressed throughout the collapse.  The expression of $Q_{\rm{LW}}$ is a fit from STARBURST99 (which used a Salpeter IMF in the range $m_{\rm{low}},m_{\rm{up}}=1,100\,  \rm{M}_{\rm{\odot}}$, an absolute metallicity of $Z=10^{-3}$ (0.05 $\rm{Z_{\odot}}$), and a stellar mass of $10^{5} \,\rm{M}_{\rm{\odot}} $).
The mean LW luminosity density $\left< L_{\rm{LW}}(M,t) \right>$ is a function of the mean number of LW photons (given by the mean production rate of LW photons per solar masses times the stellar mass of the halo), their energy and the escape fraction of these photons (we assume $f_{\rm{esc}}=1$ in this study to be able to compare with the fiducial model of D14):
\begin{equation}
\left< L_{\rm{LW}}(M,t) \right>=\frac{h \left< \nu \right>}{\Delta \nu } \left<Q_{\rm{LW}} (t) \right> f_{\rm{esc,LW}} \left( \frac{M_{\rm{\star}}}{M_{\rm{\odot}}} \right).
\end{equation}
The flux at a distance $r$ then becomes:
\begin{equation}
\left< J_{\rm{LW}}(r,M,t_{\rm{ff}}) \right> = \frac{1}{4 \pi} \frac{\left< L_{\rm{LW}}(M,t) \right>}{4 \pi r^{2}} f_{\rm{mod}}(r),
\end{equation}
where the first factor $1/4\pi$ is required to express $\left< J_{\rm{LW}}(r,M,t_{\rm{ff}}) \right>$ in $J_{21}$ units ($\rm{erg \, s^{-1} \,cm^{-2}\, Hz^{-1}\, sr^{-1}}$).
 $ f_{\rm{mod}}(r)$ is used to correct the radiation intensity for the extra dimming introduced by the LW horizon \citep{Ahn2008}:

  \begin{align}
f_{\rm{mod}} (r)&=  1.7 \exp\left(-\left(\frac{r_{cMpc}}{116.29 \alpha}\right)^{0.68}\right)-0.7 &&  \rm{if}~ r_{\rm{cMpc}}/\alpha \le 97.39\\
&=  0 &&   \rm{otherwise}.
\end{align}
 
 \noindent
Where the size is expressed in comoving Mpc (cMpc). We assume in our study that each halo has a 10\% probability of being star forming for all redshifts, $P_{\rm{SF}}=0.1$,  in agreement with \citet{Dijkstra2008}. Therefore, only 10\% of the halos in the box are considered to compute the radiation intensity, and only 10\% of nearby halos contribute to this radiation intensity. Halos are chosen randomly. The experiment is repeated 40 times to take into account the random choice of halos which are flagged as star-forming. 

\begin{figure}
	\centering
	\resizebox{\hsize}{!}{\includegraphics{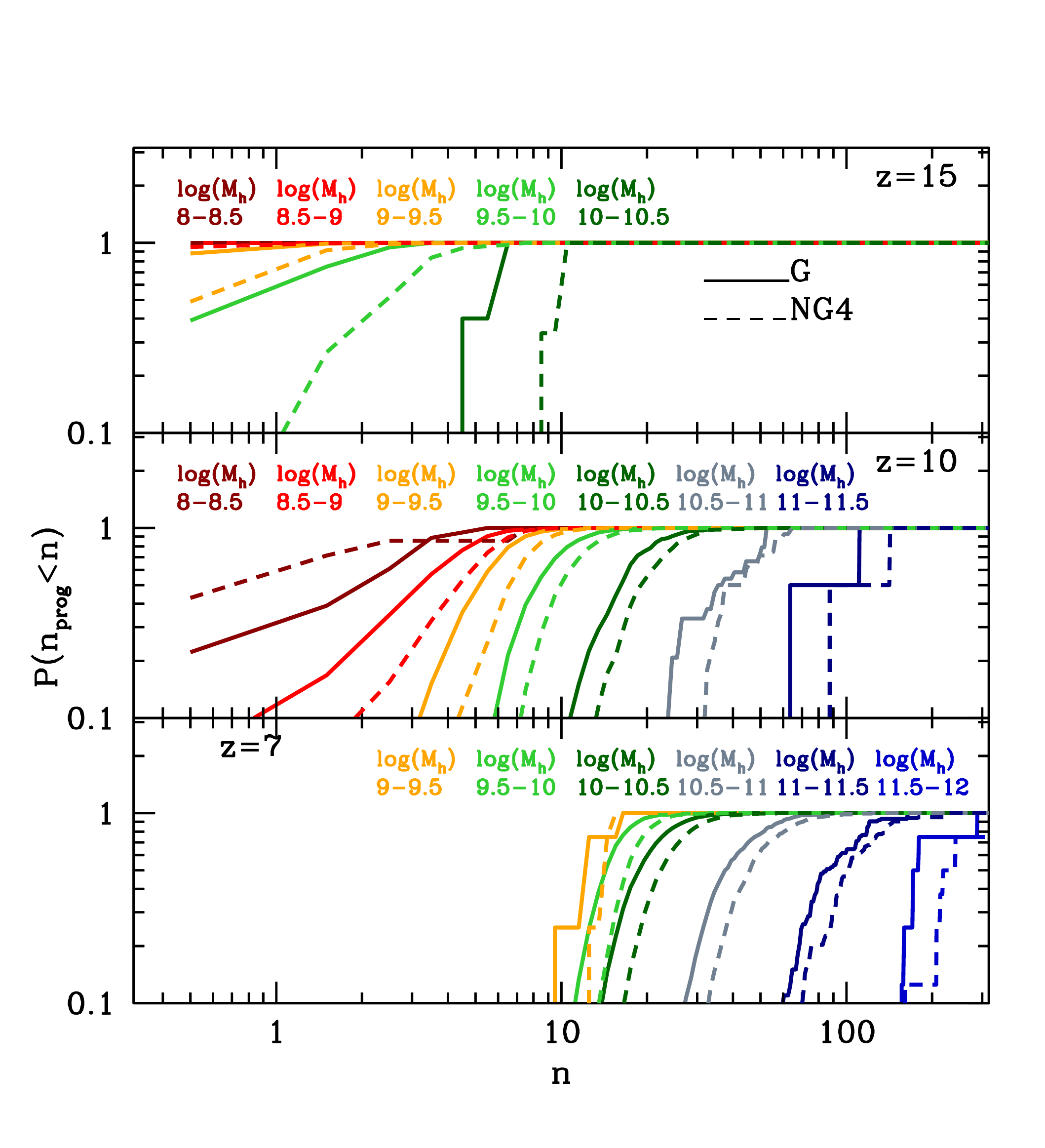}}
	\caption{Probability of having given number of progenitors for halos in a given mass range and at a given redshift ($z=15$, $z=10$, $z=7$ from the top to the bottom panel). The Gaussian simulation is represented with lines, whereas dashed lines are for the non-Gaussian simulation NG4. The typical number of progenitors is larger for NG4 at almost all masses and redshifts.}
	\label{fig:progng4}
\end{figure}

Two main opposite factors  influence the number of potential eligible DC regions: the LW radiation intensity coming from nearby star-forming regions  illuminating halos and the metal pollution they can be exposed to.  Halos irradiated by the LW flux coming from a nearby star forming halo can also be polluted by metals released at the end of the lives of  the same stars which produce the radiation. Halos that are metal-enriched would be able to cool too efficiently to be potential DC regions anymore.
The metal pollution of a halo can come from three different contributions:
(i) the contamination by the halo itself if it is star-forming, (ii) the contamination from the past history of the halo, and (iii) the potential contamination by close star-forming regions because of SN-driven galactic winds which spread metals in their surroundings. \\

To account for the first source of pollution (i), we eliminate from the list of potential DC candidates the halos which are star-forming at the current time, with the probability of being star-forming $P_{\rm{SF}}=0.1$ as described above. To account for the second contamination (ii), we estimate the probability that a halo had a progenitor which was star-forming in the past. In a hierarchical theory of structure formation, halos are formed through the continuous merging of smaller structures, which may have already encountered supernova-driven metal-enrichment episodes, making the present halo metal-polluted. Therefore the probability for a halo to be metal-polluted increases with the number of their progenitors. For a halo of a given mass and at a given redshift, the number of progenitors is on average larger for NG4 (dashed curves), than for G. For instance, at $z=15$ halos with mass $10^{10}-3.16\times 10^{10} \, \rm{M_{\odot}}$ have a 50\% probability of having less than 5 progenitors in G, and a 50\% probability of having less than 10 progenitors in NG4. In order to account for this effect we compute the mean number of progenitors per halo, for different halo mass bins, and redshifts, shown in Fig.~\ref{fig:progng4}. The mean number of progenitors is derived from the merger trees described in section 2.2.

The probability for a halo to be metal-polluted by heritage, i.e. to have metal-polluted progenitors $P_{\rm{SF \, \,progenitor}}\vert_{\rm{M_{\rm{h}},\rm{z}}}$ is described by:
\begin{equation}
P_{\rm{SF \, \, progenitor}}\vert_{\rm{M_{\rm{h}},\rm{z}}}= P_{\rm{SF}} \times \left< \rm{number\, of \,progenitors}\right >\vert_{\rm{M_{\rm{h}},\rm{z}}}. 
\label{eq:progproba}
\end{equation} 

We keep as potential DC candidates only those halos which, after Monte Carlo sampling this probability, result metal-free.

\begin{figure}
	\centering
	\resizebox{\hsize}{!}{\includegraphics{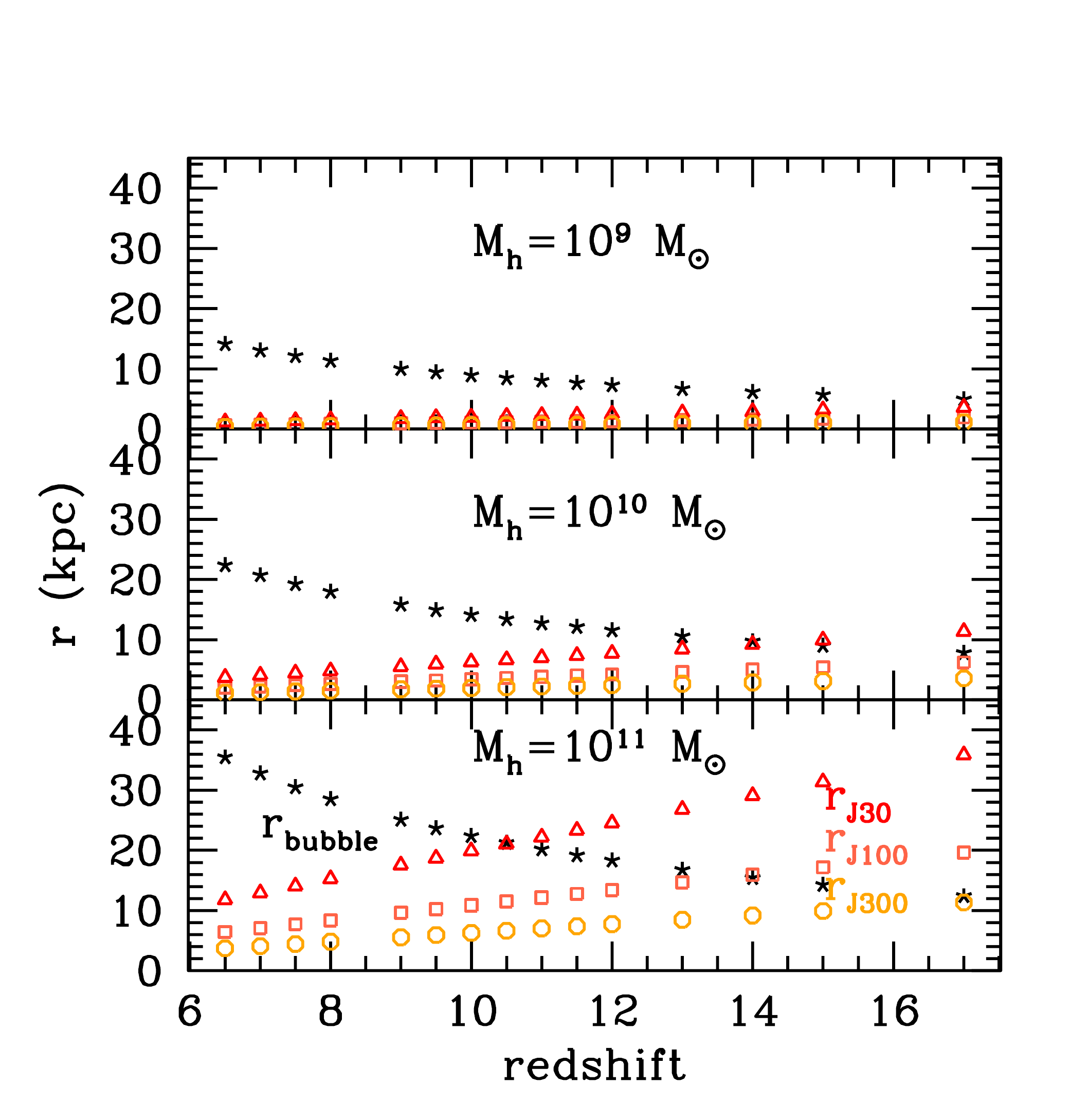}}
	\caption{Metal polluted bubble radius (black stars), and radius of the regions where $J_{\rm{21,LW}}=30$ (red triangle), $J_{\rm{21,LW}}=100$ (orange squares) or $J_{\rm{21,LW}}=300$ (yellow circles) vs redshift for different halo masses ($10^{9}, 10^{10}, 10^{11} \rm{M_{\odot}}$). All quantities are computed one free-fall time after the star-formation burst. Only regions which are at a distance above the distance given by $r_{bubble}$, and below the $r_{rad}$ are illuminated by the given radiation intensity and are not polluted by galactic winds. For instance, halos with mass $10^{11} \rm{M_{\odot}}$ at $z=15$ can irradiate a nearby halo at a distance of $\sim 17$ kpc with an intensity $J_{\rm{21,LW}}=100$ without polluting it (the metal bubble has reached only a distance of $\sim 14$ kpc).}
	\label{fig:bubble}
\end{figure}

Regarding the last source of metal pollution (iii), D14 conclude that metal pollution from nearby galaxies, through galactic winds, could be an important aspect of the halo candidates contamination.  Including the redshift dependence of density and free-fall time in the expression provided by D14 for the bubble radius, in proper kpc (pkpc), of a metal polluted bubble one free-fall time after the SF burst:
\begin{equation}
r_{\rm bubble}=22\, {\rm pkpc} \left(\frac{M_{\rm{h}}}{10^{11}\,M_{\rm{\odot}}} \right)^{1/5} \left(\frac{1+z}{11}\right)^{-6/5},
\label{eq:eqbubble}
\end{equation}
while the radius $r_{\rm rad}$ of the sphere where $J_{\rm{21,LW}}=100$  one free-fall time after the star-formation burst scales as:
\begin{equation}
\begin{split}
r_{\rm rad}=&126 \, {\rm pkpc} \times \left( \left( 1+\frac{83}{4}\left(\frac{1+z}{11}\right)^{-3/2}\right)^{-3/2} \exp \left(-\frac{83}{300}\left(\frac{1+z}{11}\right)^{-3/2} \right) \right)^{1/2} \\
&\times \left(\frac{M_{\rm{h}}}{10^{11}M_{\rm{\odot}}}\right)^{1/2}  \left(\frac{J_{\rm{21,LW}}}{100}\right)^{-1/2} \left(\frac{f_{\rm{mod}}}{1}\right)^{1/2}.
\end{split}
\end{equation}
Fig.~\ref{fig:bubble} compares the radius of the metal polluted sphere ($r_{\rm bubble}$) to the sphere ($r_{\rm rad}$) where $J_{\rm{21,LW}}=100$  or $J_{\rm{21,LW}}=300$.  

A correction that accounts for galactic winds coming from nearby star-forming galaxies is then added: if the distance between the halo we are considering as a DC candidate and a SF halo is less than $r_{\rm bubble}$, then the candidate halo would be metal-polluted, hence not an eligible DC region anymore. Fig.~\ref{fig:bubble}, however, shows that only halos with  mass $\sim10^{11} M_{\rm{\odot}}$ at $z>13$ can act as catalysts of a DC process in a nearby halo if $J_{\rm{21,LW,crit}} \leqslant 100$. At lower masses and redshift the metal polluted bubble is always larger than the bubble irradiated by sufficiently high UV flux. In our simulations,  we do not have any halos with mass $>10^{11} M_{\rm{\odot}}$ at $z \ge 11$ or $>10^{10} M_{\rm{\odot}}$ at $z>16$, and for lower-mass halos at lower redshift, as shown in Fig.~\ref{fig:bubble}, the bubble size is larger than the sphere irradiated by $J_{\rm{21,LW}}=100$  or $J_{\rm{21,LW}}=300$. Adding this correction, therefore, would leave no DC candidate in the simulation box.

\subsection{Results}   
\subsubsection{Number density of potential DC  regions}

\begin{figure}
	\centering
	\resizebox{1.0\hsize}{!}{\includegraphics{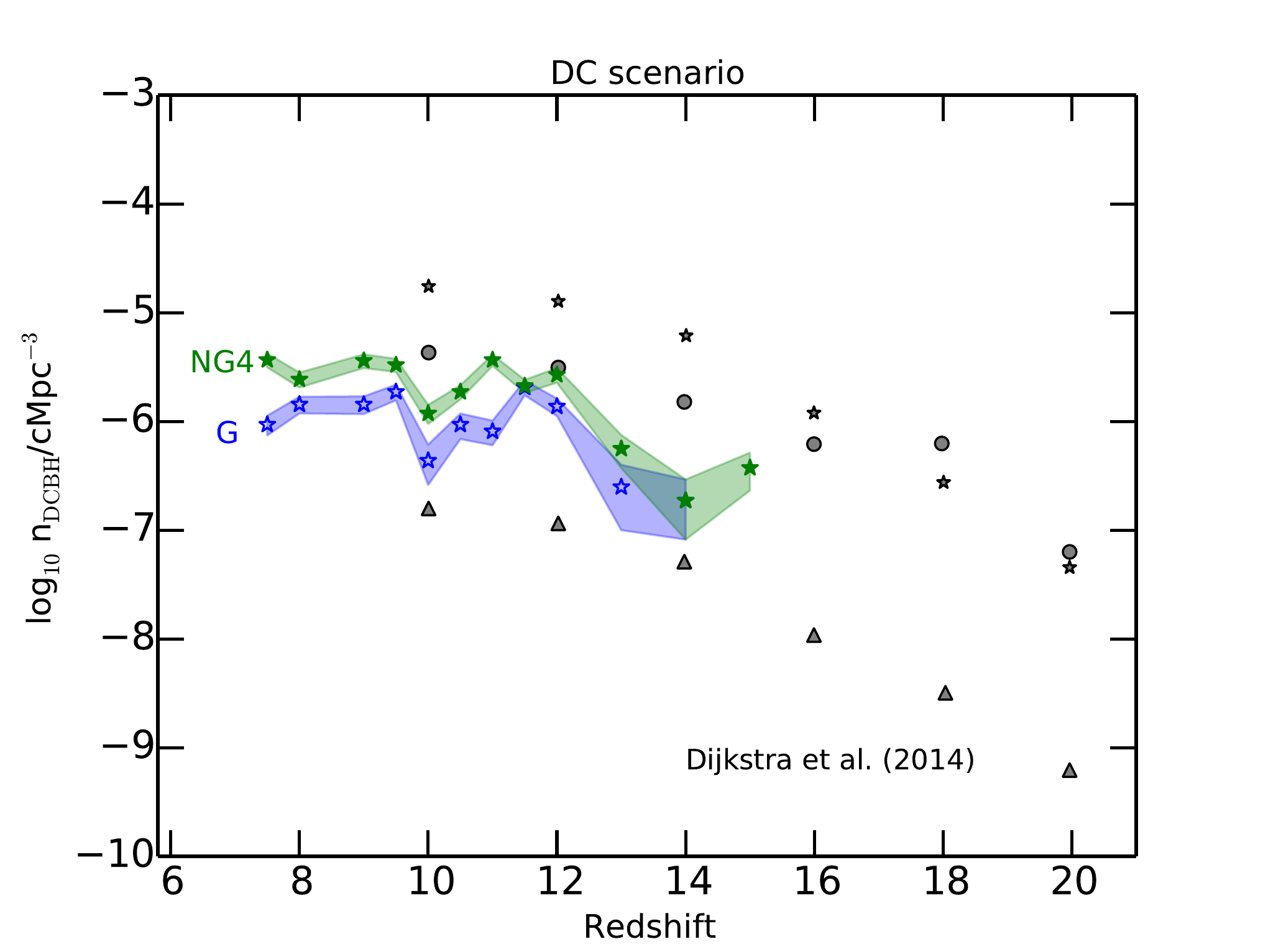}}
	\caption{Number density of DC regions identified at a given redshift in the Gaussian (blue star symbols) and non-Gaussian (green star symbols) simulations. Blue and green stars are derived from a model which does not account for direct pollution by galactic outflows, and where we use the radiation intensity threshold $J_{\rm{21,LW,crit}}=300$. Shaded areas represent the Poissonian errorbars derived from 40 realizations of the process.
The D14 results are shown in grey symbols: triangles correspond to their fiducial model where $J_{\rm{21,LW,crit}}=300$ and account for galactic winds pollution, circles to $J_{\rm{21,LW,crit}}=100$, and stars to $J_{\rm{21,LW,crit}}=300$ without considering galactic winds pollution.
Blue and green star symbols in our study can be compared with star symbols in D14 as they use the same modeling for the radiation intensity (the only differences being the probability of genetic pollution, and the use of an analytical model versus a cosmological simulation). Errorbars represent the uncertainty of the mean value of the number density of BHs.}
	\label{fig:nDCBH}
\end{figure} 

The number density of DC regions obtained in this study is shown in Fig.~\ref{fig:nDCBH}, for both G and NG4. Blue star symbols represent the number density of DC regions in G and green star symbols in NG4, using a model where we consider $J_{\rm{21,LW,crit}}=300$. 

\begin{figure}
	\centering
	\resizebox{1.0\hsize}{!}{\includegraphics{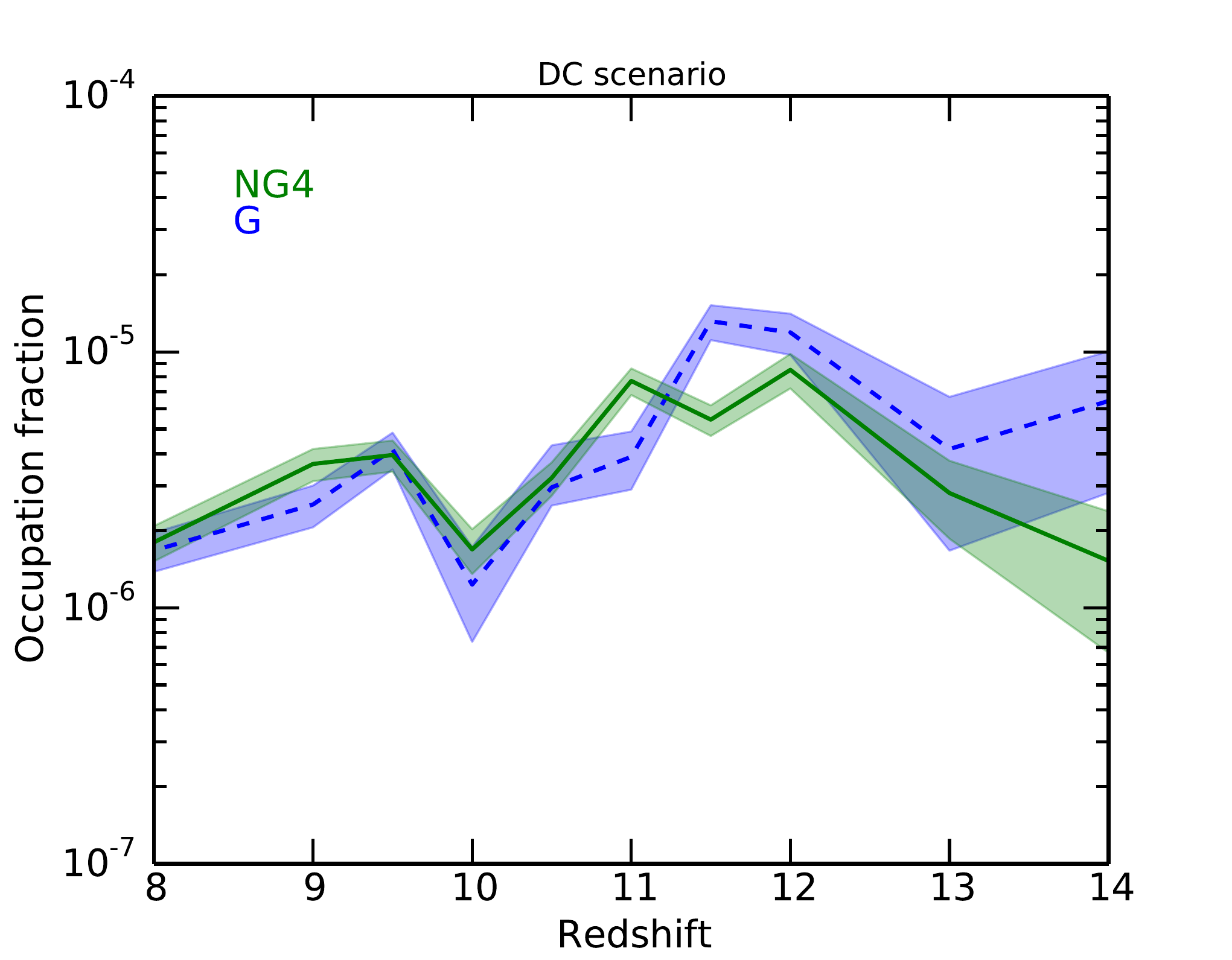}}
	\caption{Halo occupation fraction of newly formed BHs for the Gaussian (dashed blue line) and non-Gaussian (solid green line) simulations, as a function of redshift, for the DC scenario (without taking into account the metal-pollution from galactic winds). This is not a cumulative probability, but the probability that a BH forms in a halo at a given redshift. Errorbars represent the uncertainty of the mean value of the occupation fraction.}
	\label{fig:occfraction}
\end{figure}

In Fig.~\ref{fig:nDCBH} we also reproduce the results of the three main models used in D14: triangles correspond to their fiducial model where $J_{\rm{21,LW,crit}}=300$ and galactic wind pollution is included, circles to a model with $J_{\rm{21,LW,crit}}=100$ and galactic wind pollution, and stars to a model with $J_{\rm{21,LW,crit}}=300$ without considering galactic wind pollution.  Star symbols in our study and in D14 can be directly compared  as they use the same modeling for the radiation intensity. The two differences between the two studies are that we use a dark matter simulation to obtain the spatial distribution of halos, rather than an analytical prescription, and that we have derived the probability for a halo to be metal-free from the mean number of progenitors (from the merger tree history) in halo mass and redshift bins, whereas D14 use an analytical prescription. Despite these differences, our study is in good agreement with D14. 

It is worth noting that our model does not include a treatment for galactic wind pollution (at the current time or in the past). If we included these effects, as discussed in section 3.1, we would not identify any DC regions in our simulation boxes, in either G or NG4.  Indeed, if we estimate the number of DC candidates  $N_{\rm{DCBH}}$ in our simulation box from the model by D14, which includes galactic wind pollution, we find that  $N_{\rm{DCBH}}$ is less than one ($N_{\rm{DCBH}}=n_{\rm{DCBH}}\times V_{\rm{box}}\approx 10^{-7}\times V_{\rm{box}} = 0.04$ with $V_{\rm{box}}$ the simulation box volume). 


With our model, we find that NG4 (green star symbols on Fig.~\ref{fig:nDCBH}) host a number density  of DC regions slightly larger than the Gaussian simulation (blue star symbols) for the whole range of redshifts, although the differences at the largest redshifts are within the $1-\sigma$ uncertainty.  The cumulative number density of BHs at redshift z=7.5 is $1.1 \times 10^{-5}$ $\rm{cMpc^{-3}}$ for G, while in NG4 the cumulative number density is almost twice, with  $2.3\times 10^{-5}$ $\rm{cMpc^{-3}}$.  The cumulative number densities of the two simulations differ by more than $1-\sigma$.  

While the number density of BHs in NG4 is larger, so is the number of halos.  In fact, when we estimate the occupation fraction of newly formed BHs, i.e., the fraction of halos as a function of redshift where a BH is potentially formed (Fig.~\ref{fig:occfraction}, this occupation fraction is not cumulative, i.e., we calculate it for newly formed BHs only) we find that the probability of a halo being seeded with a BH is almost identical in the two simulations, although at the highest redshifts the occupation fraction in the Gaussian case is slightly above the non-Gaussian one. This can be explained as follows: since the number of progenitors is larger in the non-Gaussian simulation (see Fig. \ref{fig:progng4}), halos in the non-Gaussian simulation have a higher probability of being metal polluted because of heritage pollution.  In summary, scale-dependent non-gaussianities boost the overall number of potential DC BHs in the Universe, but not the probability that a halo hosts or not a BH.
 
\subsubsection{In the vicinity of the two most massive halos}
The analysis presented in the previous section highlights the difficulty of finding a significant number of DC regions. In order to have a clearer picture of the interplay between irradiation and metal-pollution in the model by D14, we focus here on the halos neighbouring the two most massive halos in our simulation volume. The reason for this choice is that according to the model described in section 3.1, only halos more massive than  $10^{11} \rm{M_{\odot}}$ can provide an intensity higher than $J_{\rm{21,LW}}=300$ at a distance of 10 pkpc one free-fall time after the star formation burst. Critically, a common critical intensity value suggested by simulations \citep{Bromm2003,Sugimura14,Latif2015, Regan2014} is  $J_{\rm{21,LW,crit}}\sim10^{3}$. 

We therefore select the two halos more massive than $10^{11} \rm{M_{\odot}}$ at redshift $z=10$ in the G and NG4 simulations. These two halos match one another in the two simulations. Only these two halos are able to produce sufficient radiation to efficiently dissociate molecular hydrogen on $\sim$ 20 pkpc distances. We consider all the halos inside a 20 pkpc radius centred on each of the most massive halos and compute the radiation intensity illuminating them.  \\     

\begin{figure}
	\centering
	\resizebox{1.1\hsize}{!}{\includegraphics{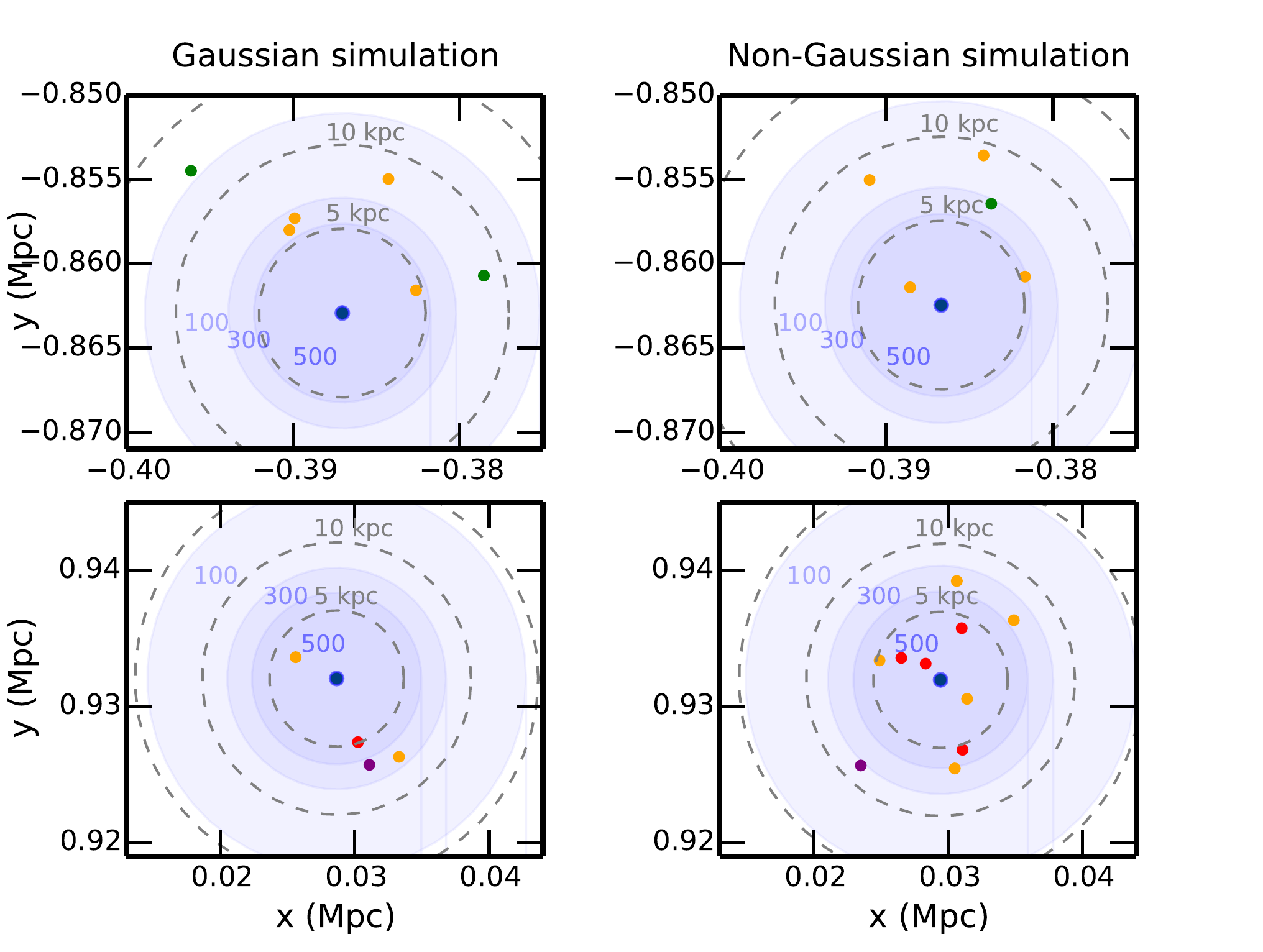}}
	\caption{The two most massive halos of the Gaussian and non-Gaussian simulations at redshift z=10 are presented in blue dots. On the top left panel, we show a halo with a mass of $1.17\times 10^{11}$ (ID 54335), on the top right $1.65 \times10^{11}$ (ID 61371), on the bottom left  $1.22 \times 10^{11}$ (ID 104966) and no the bottom right $1.75 \times 10^{11} M_{\odot}$ (ID 118759).  Indicative radii of 5 kpc, 10 kpc, and 15 kpc are shown with grey dashed lines in the ($x,y$) plane. The radiation intensity from these massive halos is shown in blue contours, the innermost area has intensity higher than $J_{\rm{21,LW}}=500$, the second by  $J_{\rm{21,LW}}=300$ and outermost $J_{\rm{21,LW}}=100$. Finally, the projection of  halos in the plane ($x,y$) is shown in colours indicating the radiation intensity they experience (in 3D): in green $J_{\rm{21,LW}}<100$, in orange  $J_{\rm{21,LW}}\geq100$, in purple $J_{\rm{21,LW}}\geq300$, in red  $J_{\rm{21,LW}}\geq500$.}
	\label{fig:massive_halos}
\end{figure} 

In Fig.~\ref{fig:massive_halos}, halos shown in red are illuminated by a radiation intensity higher than $J_{\rm{21,LW}}=500$. The number of halos is higher for the non-Gaussian simulation, as well as the number of halos exposed to a high radiation intensity. In NG4, which forms more low-mass halos, the potential number of DC regions is increased.  However, if we account for SN-driven metal-pollution using Eq.~\ref{eq:eqbubble}, 1 Myr after the SN explosion, the metals in the massive halo are already spread over 4 pkpc. After 2 Myr, the metal-polluted sphere reaches 5-6 pkpc. At this time all halos illuminated by a LW intensity $J_{\rm{21,LW}}>300$ are inside this sphere and therefore polluted by metals, making the DC process unfeasible.  \\
Within the formalism we have adopted here, we can not identify a difference between G and NG4. However, this model includes several simplifications, for instance the expansion of the metal bubble in a real Universe may not be spherical, and  $P_{\rm{SF}}$ may well be a function of redshift and halo mass. We argue that the non-Gaussian simulation, having more low-mass halos irradiated by a strong UV flux, could represent a more favourable environment for this scenario.

\section{BHs formed from the remnants of the first generation of stars.}
Pop III star remnants is another  popular scenario to explain the formation of BH seeds in the early Universe \citep{Madau2001,Volonteri2003}. BHs are predicted to form in metal-free mini-halos ($\rm{M_{h}} \sim 10^{5}$ \Msun) at redshift $z=20-30$ from the remnants of the first generation of stars (the so-called PopIII stars, which are stars without  elements heavier than hydrogen and helium). These stars have never been observed so far, nevertheless they are thought to have masses ranging from $10$ to $1000$ \Msun $\,$ \citep[][and references therein]{Bromm_Yoshida_2011,2015MNRAS.448..568H}. If some of these stars are sufficiently massive ( $>$ 260 \Msun), BHs retaining up to half the stellar mass are formed, leading to the formation of a BH seed of $\sim$100 \Msun  $\,$\citep{2001ApJ...550..372F}. \\
In this section we want to estimate the number density of BHs and the  fraction of halos where a BH can form via the Pop III stars scenario for G and NG4. We stress that our simulations do not have the resolution needed to resolve mini-halos, therefore the following experiment can only be used to assess trends. However, since the model we consider in this work has stronger non-Gaussianity on smaller scales (and thus on less massive halos), we can expect that the impact of non-Gaussianities on mini-halos can be even larger than what we find (in the following paragraphs) for more massive halos. \\
According to the PopIII star scenario, only metal-free halos can host the first generation of stars. We therefore identify all the star-forming and metal-free halos in the two simulations using the same approach described in section 3.1. The probability of a halo being star-forming is again $P_{\rm{SF}}=0.1$ , identical for all redshifts, meaning that only 10\% of the halos are selected in the first place as potential hosts of a Pop III remnant BH seed. Additionally, we ensure that these halos are not metal-polluted from the past history of the halo (heritage pollution), nor from galactic winds coming from neighboring halos at a coeval redshift.\\
Regarding the second aspect, we account for the probability of having a star-forming neighbor $P_{\rm{SF}}=0.1$ on a distance scale $r_{\rm{bubble}}$ defined in  Eq. ~\ref{eq:eqbubble}, this distance is redshift and halo mass dependent. We also consider the probability for the neighboring halos (on the same distance scale) to have spread metals in their past history, which could also have introduced metals in the considered halo, making it ineligible to form PopIII star in a metal-free environment. We perform 40 realizations of the model.

\begin{figure}
	\centering
	\resizebox{\hsize}{!}{\includegraphics{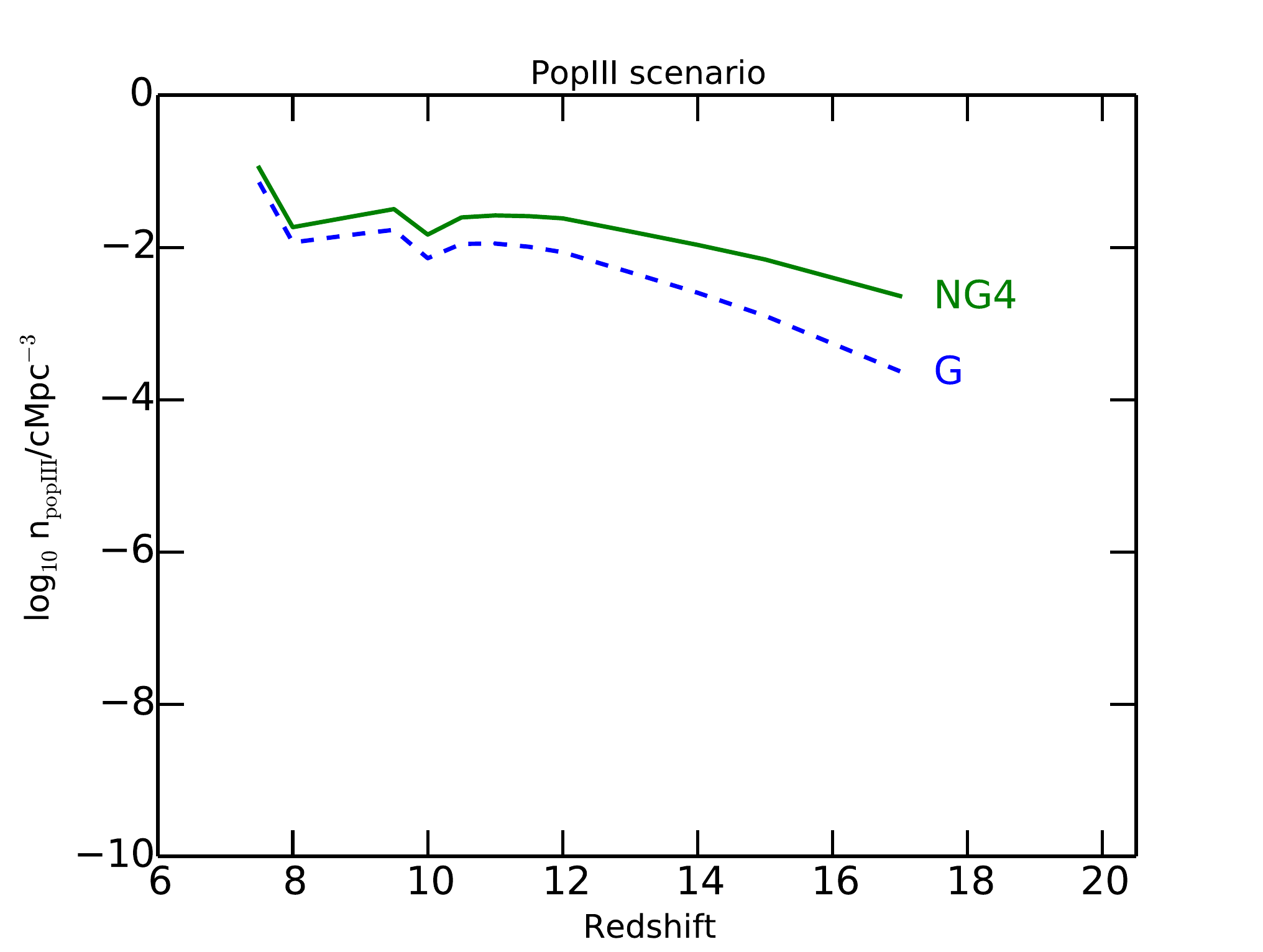}}
	\caption{Number density of Pop III star remnant BHs formed at a given redshift for the Gaussian (dashed blue line) and the non-Gaussian (solid green line) simulations. Errorbars represent the uncertainty of the mean value of the number density, the uncertainty is here too small to be seen.}
	\label{fig:densitypopIII}
\end{figure} 

\begin{figure}
	\centering
	\resizebox{\hsize}{!}{\includegraphics{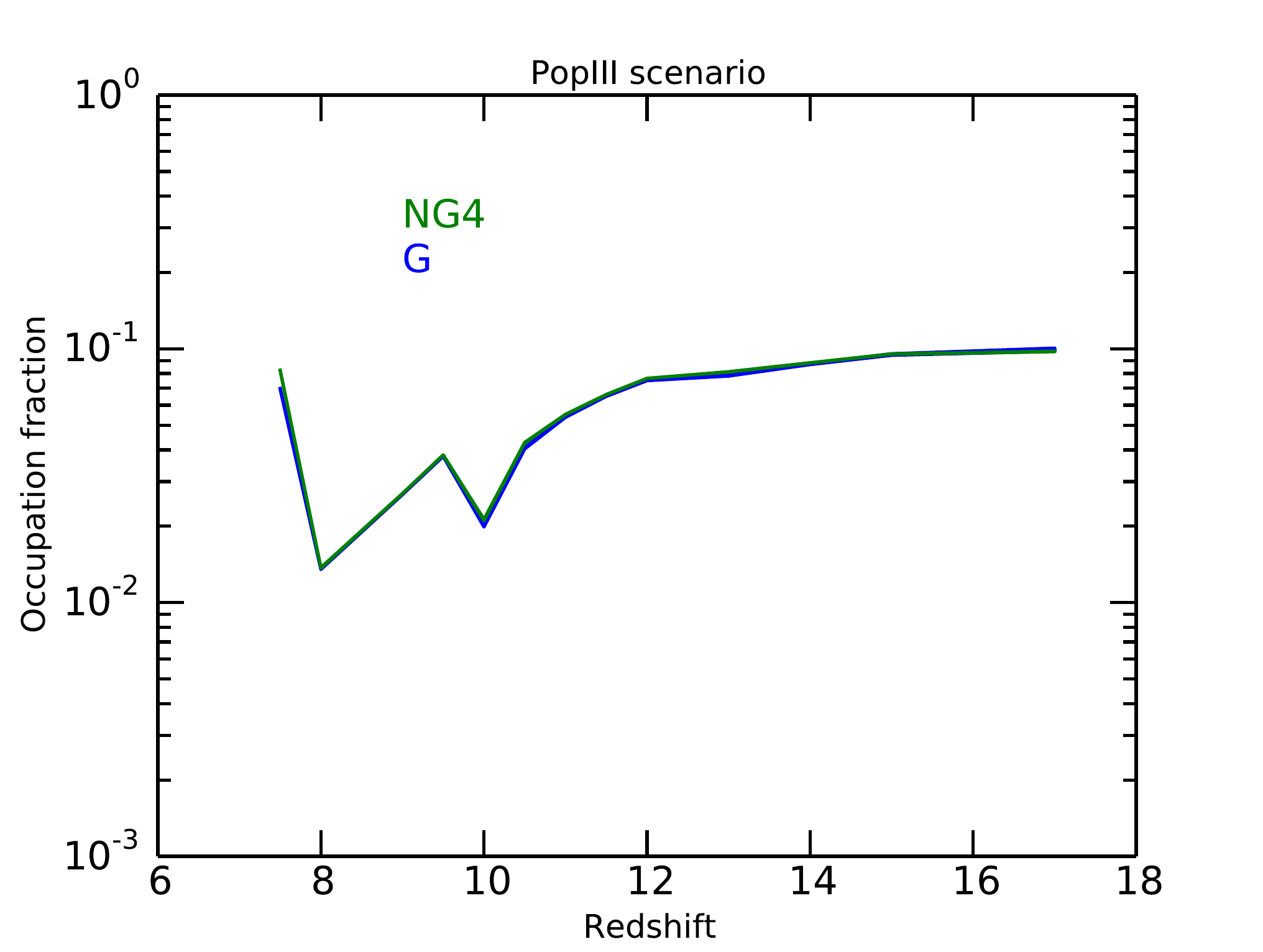}}
	\caption{Newly formed BH-halo occupation fraction for the Gaussian (blue line) and the non-Gaussian (green line) simulations, as a function of redshift, for the remnants of the first generation of stars scenario. As in Fig.~\ref{fig:occfraction}, this is not a cumulative probability. Errorbars represent the uncertainty of the mean value of the occupation fraction, the uncertainty is here too small to be seen.}
	\label{fig:occfractionpopIII}
\end{figure}

Fig. \ref{fig:densitypopIII} represents the mean number density of potential BHs formed via the Pop III star remnant scenario for the two simulations (Gaussian in blue, non-Gaussian in green). The trends of the two curves are similar, but NG4 hosts more BHs. The enhancement in the number density of BHs increases with redshift, while at  $z=7.5$  the two curves are almost overlapping. However the cumulative number density of BHs for NG4 is again almost twice as large (G: 0.17 $\rm{cMpc}^{-3}$, NG4: 0.34 $\rm{cMpc}^{-3}$).  The cumulative number density in the two cases differs by more than $1-\sigma$.
The occupation fraction of halos where BHs form via this scenario is shown in Fig.~\ref{fig:occfractionpopIII}. The blue line indicates the occupation fraction for G, and the green line NG4. We note that the occupation fraction is almost identical for the two simulations: it is $\sim 10^{-1}$ at  $z=17$ and drops to $\sim 10^{-2}$ at  $z=8$ before increasing again (we account for forming BHs only, not the cumulative occupation fraction). The Gaussian case is slightly above the non-Gaussian one. This can be explained with the same arguments as those discussed for the DC case, and, moreover, halos in NG4 have also a number of neighbours slightly higher than halos in G, increasing the probability of being polluted by galactic winds. 
As noted above, our simulations do not resolve mini-halos, but since our model for $f_\rmn{NL}$ enhances the number of low-mass halos at a given redshift, there will be more mini-halos in the non-Gaussian case, favouring the formation of PopIII stars at even higher redshifts than those considered here.  Therefore, also at higher redshift, the number of BHs formed throughout the PopIII remnant scenario would be higher in the non-Gaussian case until metal pollution starts dominating the environment. Our results can therefore be considered a lower limit to the enhancement in the BH population.

\section{BHs in the most massive halos at  $z=6.5$}
We now turn to exploring the possibility that the different growth histories of halos in Gaussian and non-Gaussian models affect the assembly of BHs at the high mass end. Using merger trees made with {\sc TreeMaker}, we derive the history of the most massive halos in all simulation boxes at redshift $z=6.5$. From the mass evolution of these halos, we derive the evolution that a hypothetical BH in these halos could have. 

To probe the {\it cumulative} effect that a different early evolution has on the BH population, we evolve the BH masses in the merger trees. Rather than assigning a BH mass simply based on the halo mass at a given time we seed the highest redshift progenitor halos of the $z=6.5$ halos with BH `seeds' and evolve their mass over cosmic time adopting simple prescriptions. Our goal is to explore how the dominant differences in halo growth histories caused by non-Gaussian initial conditions affect the assembly of the BHs. The main diagnostics will be the mean BH mass as a function of time and the number of BHs with mass above some minimum threshold. The latter diagnostic is important as we are currently able to detect only the most massive BHs ($\sim10^9$ \Msun). Even in the future, at such high redshift, we will always pick the most massive BHs, although the mass threshold will decrease. For instance, the future X-ray mission ATHENA\footnote{http://sci.esa.int/cosmic-vision/54517-athena/} is expected to be able to detect BHs with masses above $10^6-10^7$ $\rm{M}_{\odot}$ up to $z\sim 8-10$ \citep{2013arXiv1306.2325A}. 

\begin{figure}
	\centering
	\resizebox{\hsize}{!}{\includegraphics{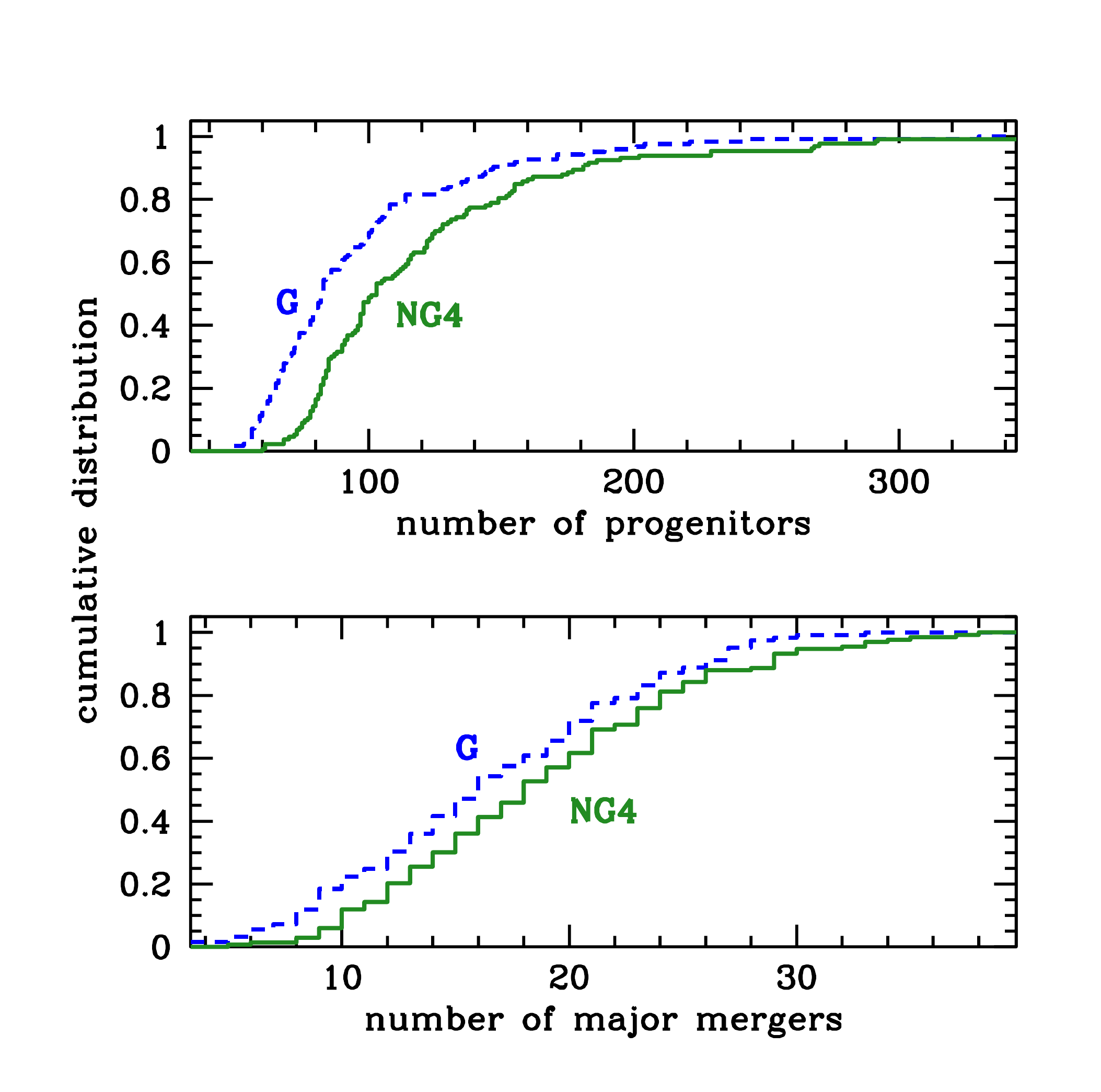}}
	\caption{Cumulative distribution of the number of progenitors (top) and halo mergers with mass ratio $>0.1$ for all halos with mass  $>10^{11} \, \rm{M}_{\odot}$ at $z=6.5$ in the Gaussian (G, dotted blue histogram) and the non-Gaussian (NG4, green, solid histogram) simulations. The probability that the progenitor distributions come from the same parent distribution is less than $10^{-6}$. The evidence for differences in the major merger distributions is weaker, 0.14.}
	\label{fig:prog_mjm_ks}
\end{figure}

Specifically, first we analyse the merger trees of all halos with mass  $>10^{11} \, \rm{M}_{\odot}$ at $z=6.5$ to find the effects of non-Gaussianities. There are 125 such halos in simulation G and 133 in NG4. The main differences we find in the halo growth are in the total number of progenitors, and in the number of mergers involving similarly sized halos, i.e., with mass ratio $>0.1$ (`major mergers' hereafter). We perform a Kolmogorov-Smirnov test (Fig.~\ref{fig:prog_mjm_ks}) and find that the probability that the progenitor number distributions come from the same parent distribution is less than $10^{-6}$. The evidence for differences in the major merger distributions is weaker, with a probability of 0.14, because of the small-number statistics. The mean number of progenitors for the $>10^{11} \, \rm{M}_{\odot}$ at $z=6.5$ halos is 95 for model G and 120 for NG4. The mean number of major mergers  is 16 (G) and 20 (NG4).

We then model, in a simplified way, the evolution of hypothetical BHs over the cosmic history of these halos. Two main factors linked to the different number of halos and progenitors in G and NG4 would influence the BH distribution (masses and number) at $z=6.5$: (i) how many halos host BHs, and (ii) the number of major mergers for merger-driven BH growth. 
Regarding the first point, it is expected that  BH formation is not ubiquitous in all halos as specific conditions are required \citep[see sections 3 and 4 and][for a review]{2010A&ARv..18..279V}. Therefore, if each halo has a given probability of hosting a BH, the larger the number of the progenitors of a halo, the higher the probability that a halo without a BH acquires a BH through a merger with a halo seeded by a BH \citep{Menou2001}. Regarding the second point, major galaxy mergers trigger torques that destabilize the gas in a galaxy, causing nuclear inflows that trigger BH accretion episodes \citep{Kauffmann2000,Hopkins2006}. If a BH is hosted in a galaxy that experiences a larger number of major mergers, its growth will be boosted. 

\begin{figure}
	\centering
	\resizebox{\hsize}{!}{\includegraphics{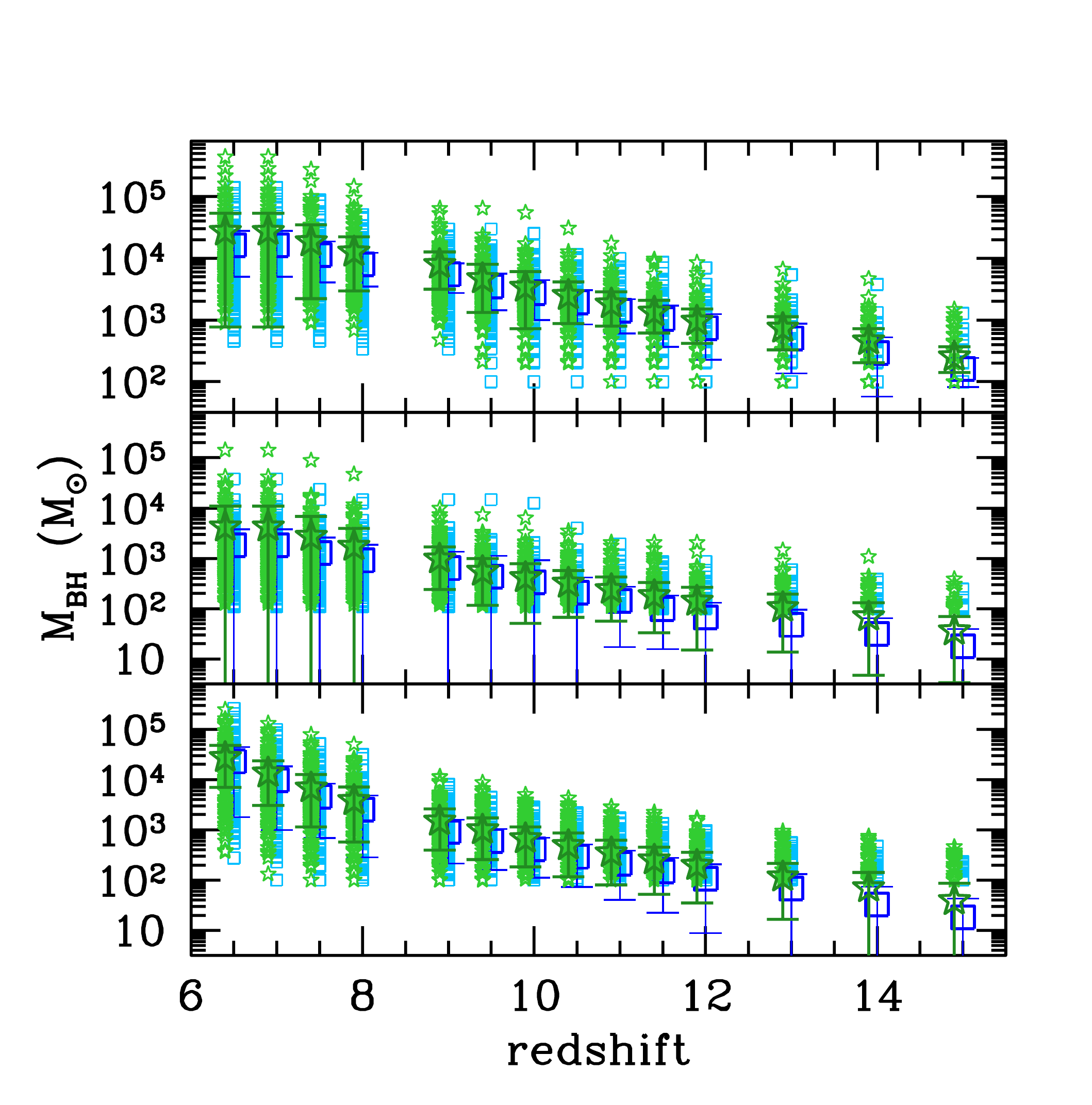}}
	\caption{Top: Evolution of the BH mass for all halos with mass $>10^{11}$ \Msun \,~at $z=6.5$, assuming that accretion is only merger-driven. 
	Middle:  assuming the probability that a halo hosts a BH is 10\%,  and accretion is only merger-driven.
	Bottom: assuming that the probability that a halo hosts a BH is 10\% and BHs grow in mass  through random accretion. 
	Simulations: G (blue asterisks); NG4 (green stars). Each halo is represented by a point at each simulation output, and we calculate mean and variance at each output redshift (shown as a larger point with errorbar).  }
	\label{fig:mjmonly}
\end{figure}

To test how the different merger histories of Gaussian and non-Gaussian models affect the BH growth in this way, we perform a first experiment where we assume that each halo starts with a $10^2$~\Msun \, BH, and, after each major merger, the BHs also accrete at the Eddington limit, assuming a radiative efficiency of 10\%, for one dynamical time  \citep{2014CQGra..31x4005T}, while the masses of the BHs in the merging halos are summed.  The results are shown in Fig.~\ref{fig:mjmonly}, top panel. NG4 has a consistently higher mean BH mass and a higher number of BHs with mass above a minimum threshold, e.g., $10^4$ \Msun \, at $z=6.5$. Simulation NG4 hosts 58 BHs with mass $>10^4$ \Msun \, at $z=6.5$, while G hosts 57. The BHs with mass $>10^5$ \Msun \, are 8 and 3 respectively. 

We perform a second experiment (Fig.~\ref{fig:mjmonly}, middle panel) where we assume that each halo has a 10\% probability of hosting a $10^2$~\Msun \, BH when it enters the merger tree. We use here the occupation fraction of the Pop III star remnant case in order to have some statistics. We note that if we increased the seed mass by a given factor, the results shown below would scale by the same factor. Given the results of sections 3 and 4, we adopt the same probability for both G and NG4.  If the main halo already hosts a BH, the masses of the BHs in the main and merging halo are summed.  BHs also accrete at the Eddington limit for one dynamical time after each major merger.  Simulation NG4 has 12 BHs with mass $>10^4$ \Msun \, at $z=6.5$, while G has 3. Above $10^5$ \Msun \, are 1 and 0 respectively. By $z=6.5$ 80\% of the halos host a BH in G, while this fraction is 90\% for NG4, despite starting with the same occupation fraction of 10\% in each case  \citep[see][]{Menou2001}.

The final experiment is to forego major-merger driven accretion and assign to each BH an accretion rate based on a distribution probability calculated in a large scale cosmological simulation, Horizon-AGN \citep{2014MNRAS.444.1453D}. In Fig.~\ref{fig:mjmonly}, bottom panel, we show a model where we assume that each halo has a 10\% probability of hosting a $10^2$ \Msun  ~BH when it enters the merger tree, and if the main halo already hosts a BH, their masses are summed.  The BHs also accrete over a timestep with an accretion rate randomly drawn from the distribution of Eddington rate, $\lambda$, calculated from all the BHs at $6<z<8$ in the Horizon-AGN simulation: $\rm{d}N/\rm{d}\log \lambda=10^{(\log \lambda+2)}/10^2$. In this case, simulation NG4 hosts 67 BHs with mass $>10^4$ Msun at $z=6.5$ while G hosts 51.  The BHs with mass $>10^5$ \Msun \, are 10 and 9 respectively. Again, at $z=6.5$ 80\% of the halos host a BH in G, while this fraction is 90\% for NG4.

The main conclusion is that in NG4 the number of the most massive BHs is larger, and the mean BH mass at $z=6.5$  increases by 0.08, 0.22 and 0.36 dex for the third, first and second experiment respectively. In the Eddington rate formalism, a mass difference of a factor of two corresponds to a change in the growth time of 70\%, because of the exponential dependence. While in all the experiments the statistical significance of the difference between G and NG4 is low (they are compatible within $1-\sigma$) the trends are always consistent: if all conditions for BH growth are equal, i.e., BH physics is the same, a population of BHs in NG4 would grow faster and have more more massive BHs. In the example shown here, however, the small high-redshift seeds do not grow much more than to a few $\times 10^5$ \Msun \, at $z=6.5$. We have tested the difference with a case where the initial seed mass is $10^5$ \Msun $\,$(keeping all other assumptions equal), and we find that, in that case, BHs can grow up to several $10^8$ \Msun, less than the masses of $z>6$ quasars. This is not surprising, given the absence, in our simulation box, of the sufficiently massive dark matter halos, $\sim 10^{13}$ $\rm{M}_{\odot}$  expected to be hosting these extremely massive BHs. To explain the observed quasars, with mass $> 10^9$ \Msun , large seeds or additional growth channels (e.g., super-Eddington accretion), and sustained accretion at the Eddington level \citep{dimatteoetal12, duboisetal12} would be needed.

\section{Discussion and conclusions}
In this paper, we have addressed the formation and the growth of supermassive BHs in the presence of  scale-dependent non-Gaussianities. We use two identical simulations except for their initial conditions, with either Gaussian or scale-dependent non-Gaussian primordial perturbations ($f_{\rmn{NL}}(k) = f_{\rmn{NL}, 0}\,\left(k/k_0\right)^{\alpha}$, with $\alpha=4/3 $ and  $f_{\rmn{NL}, 0}=10^{4}$).
The introduction of these non-Gaussianities on galactic scales, consistent at larger scales with the Planck results, produces an enhancement in the low-mass end of the halo and galaxy mass functions, increasing with redshift. As a consequence, changes in the BH population arise as well.  We explore the impact of scale-dependent non-Gaussian primordial perturbations on two models of BH formation, and on the growth of the putative BHs. 
Sherkatghanad \& Brandenberger (2015) also investigate local-type non-Gaussianities, i.e. with both skewness ($f_\rmn{NL}$) and kurtosis (described by the parameter $g_\rmn{NL}$),  in the context of BH formation. They do not include scale-dependent  non-Gaussianities, and conclude that non-Gaussianities do not strongly affect the number density of dark matter halos at high redshifts (and of BHs as a consequence). This is in agreement with our previous work \citep{Habouzit2014} where we showed that non-Gaussian models closest to a non-scale dependent $f_\rmn{NL}$ do not show significant differences in halo and stellar mass functions compared to the Gaussian model. 
On a related note, \citet{Hirano15} find that varying the slope of the primordial power spectrum impacts the formation of structures as well: an enhanced power spectrum at small length scales (or blue-tilted power spectrum) pushes to the formation of the first stars at much higher redshifts, and the higher CMB temperature leads to more massive stars, which can be precursor of massive BHs.

The formation of DC BHs is predicted to happen in metal-poor regions illuminated by a UV radiation intensity higher than a critical value (here we use $J_{\rm{21,LW,crit}}=100$).
We have implemented a model to identify these regions, inspired by D14, to compute the radiation intensity emitted from galaxies forming in dark matter halos.
The increase in the galaxy mass function, particularly at the low-mass end, in the non-Gaussian simulation leads to a larger number density of potential DC regions. This is due to the increase of the number of galaxies for two reasons, there are statistically more regions that can collapse forming a BH, but also because more galaxies can act as radiation sources to illuminate dense regions where the collapse may happen. 
Conversely, the presence of more galaxies can also lead to a stronger metal enrichment, making a halo unavailable for the DC process. This last aspect has been difficult to study: we have implemented a model for the metal-pollution coming from close star-forming regions, in the current time and the past history of the regions. Taking into account the pollution coming from galactic winds reveals a metal pollution of all the previously identified DC regions, making  any comparison between the Gaussian and non-Gaussian simulations impossible. A larger simulation box would be needed to test in further detail the impact of the enhancement in the low-mass end of the galaxy mass function on the metal-enrichment of potential DC regions by galactic winds. 

However, as the critical value for the radiation intensity is still highly debated, and may be as high as $J_{\rm{21,LW,crit}}=10^3$, only halos as massive as $10^{11} \rm{M_{\odot}}$ or larger could provide sufficiently high radiation to suppress molecular hydrogen in their neighbourhood. The number of neighbours in the vicinity of the two halos more massive than $10^{11} \rm{M_{\odot}}$ in the non-Gaussian simulation is larger, up to a factor 4, for halos seeing a radiation intensity $>J_{\rm{21,LW}}=500$ in the example shown in Fig. ~\ref{fig:massive_halos}. This illustrates the effect of primordial non-Gaussianities in increasing the number density of DC regions. Metal pollution remains, however, a concern. Two factors may alleviate the importance of metal pollution: in the first place, SN bubbles may not be spherical, as assumed in D14 and our calculation, once a realistic gas and DM distribution is taken into account. Additionally, we and D14 have assumed, following \cite{2001ApJ...555...92M} a simplified evolution of the bubble radius (see also section 5 in D14). A third approximation we and D14 have made is that the probability of a halo being star-forming is constant with redshift and halo mass. These issues will be studied in a companion paper. 

A second path for BH formation we have explored hinges on the remnants of the first generation of stars, in metal-free mini-halos. In order to test the impact of primordial density perturbations on this scenario, we have modified the scheme we have adopted for the direct collapse scenario (same probability for a halo to be star-forming, and  the same contributions for the metal-pollution, namely pollution from heritage of the considered halo itself, and from galactic winds coming from neighboring star-forming halos).  Only star-forming and metal-free halos are considered as eligible site to form BHs. While our simulations have a much lower resolution than needed to resolve mini-halos, we can at least identify some trends. We find that non-Gaussianities do not have a strong effect on the newly formed BH-halo occupation fraction, in both cases the occupation fraction drops from $10^{-1}$ at $z=20$ to $10^{-2}$ at $z=8$. Conversely, the number density of BHs is increased at the highest redshifts in the presence of non-Gaussianities, up to one order of magnitude. The larger number of progenitors and neighbours in the non-Gaussian simulation imply a larger probability for a halo to be/become polluted by metals.

The growth of supermassive BHs is also altered when considering non-Gaussianities.  After deriving the merger history of the most massive halos at $z=6.5$ in both the Gaussian and non-Gaussian simulations, we study the evolution of BHs in massive halos down to $z=6.5$.  To investigate the cumulative effect over cosmic times on the BHs assembly, we model the growth of BHs in three different ways. Different probabilities for a halo of hosting a seed BH, and different accretion models (either each BH accretes at the Eddington limit for a dynamical time after a major merger or using an accretion rate based on a distribution probability derived from a large-scale hydrodynamical simulation) are adopted.
We have not included in our models the effects of ``kicks" caused by asymmetric emission of gravitational waves, which have been proposed to be possibly responsible for ejecting BHs from halos with shallow potential wells, thus halting or reducing the growth of high-redshift BHs hosted in small halos \citep[e.g.][]{YooMiralda2004,VRees2006,2009ApJ...696.1798T}. This effect, however, seems to affect less than 10\% of binaries and it becomes negligible for BH mergers at $z<10$ \citep{VRees2006}. 
We find that  non-Gaussianities imply a larger number of massive BHs and also an increase in the mean BH mass (up to 0.36 in the most favourable experiment). 
A population of supermassive BHs will then grow faster and to higher masses in a universe with scale-dependent non-Gaussian primordial density fluctuations. If the seed masses are similar to those of PopIII star remnants, BHs will not be able to grow above few $\times 10^5$  $\rm{M}_{\odot}$ by $z=6$. However, our simulations do not resolve mini-halos, and we may underestimate the growth of seeds at earlier times. We argue that, in a simulation resolving mini-halos, BHs would have formed earlier through the PopIII remnant scenario, leading to a longer time for them to grow in mass. If we assumed that PopIII remnant seeds with mass 100 \Msun form at $z\sim30$ in halos unresolved in our simulations, they would have grown, assuming, optimistically, constant growth at the Eddington rate \citep[but see][]{JBromm,Alvarez2009,Milos2009,Park2010} to $\sim 10^3$ $\rm{M}_{\odot}$ by $z=18$, where we start our analysis. The final BH mass at $z=6$ would then be $\sim$ one order of magnitude larger, a few $\times 10^6$ \Msun, still short of the $\sim 10^9$ $\rm{M}_{\odot}$ required. The very limited growth obtained for the PopIII remnant case suggests that large seeds or super-Eddington accretion \citep[see][and references therein]{2015ApJ...804..148V} may be necessary for successful BH growth. We have done the same experiments on BH growth starting with initial $10^5$ $\rm{M}_{\odot}$ BH masses (not shown in the paper, but see section 4). In this case we found that it is much easier for BHs to grow to higher BH masses, but still only to several $10^8$ \Msun. This is not unexpected, because our simulation box does not contain the very rare and biased dark matter halos with masses$\sim 10^{13}$ $\rm{M}_{\odot}$ believed to be hosting these extreme BHs. 

\section*{Acknowledgements} 
We acknowledge funding support for this research from a Marie Curie FP7-Reintegration-Grants  (PCIG10-GA-2011-303609, MV and MH) and from the European Research Council (FP7/2007-2013 Grant Agreement no.\ 614199, project ``BLACK'', MV and ML) within the 7th European Community Framework Programme. 
The simulations were performed on Horizon, cluster of the Institut d'Astrophysique de Paris, we acknowledge Stephane Rouberol for technical support on this work.
\bibliography{biblio}

\label{lastpage}
\end{document}